\documentclass[twocolumn,showpacs,preprintnumbers,amsmath,amssymb,prl]{revtex4} 
 
\usepackage{graphicx}
\usepackage{dcolumn}
\usepackage{bm}
 
\newcommand{\epsfigbox}[5]{%
\begin{figure} \vspace{#3}%
\includegraphics{#2}%
\caption{ 
\label{fig:#1} #5} 
\vspace{#4} 
\end{figure}}

\newcommand{\quarterthin}{\kern 0.0417em}

\newcommand{\ket}[1]{|#1\rangle}

\begin{document} 
 
 
 
\title{ 
Mott Insulators, No-Double-Occupancy, and Non-Abelian Superconductivity} 
 
\author{ 
Mike Guidry$^{(1)}$,  
Yang Sun$^{(2)}$, and 
Cheng-Li Wu$^{(3,1)}$ 
}

\affiliation{ 
$^{(1)}$Department of Physics and Astronomy, University of Tennessee,  
Knoxville, Tennessee 37996, USA \\ 
$^{(2)}$Department of Physics, University of Notre Dame, Notre Dame, Indiana 46556, USA \\ 
$^{(3)}$Physics Division, National Center for Theoretical Science,  
Hsinchu, Taiwan 300, ROC} 
 
\begin{abstract} 
$SU(4)$ dynamical symmetry  
is shown to imply a 
no-double-occupancy constraint on the minimal symmetry description of 
antiferromagnetism and $d$-wave superconductivity.  This implies  
a maximum doping fraction of 
$\tfrac14$ for cuprates and provides 
a microscopic critique of the projected SO(5) model.  
We propose that $SU(4)$ superconductors  
are representative of a class of compounds that 
we term {\em non-abelian superconductors}. 
We further 
suggest that 
non-abelian superconductors may exist 
having $SU(4)$ symmetry and therefore cuprate-like dynamics, but 
without $d$-wave hybridization. 
\end{abstract} 
 
\pacs{74.20.-z, 74.72.-h, 11.30.-j, 02.20.-a} 
\date{\today} 
\maketitle 
 
 
 
Suppression of double occupancy on sites in 
the copper oxide planes  
is critical in explaining 
why cuprate systems are antiferromagnetic  
Mott insulators at half filling and become superconductors through  
hole doping.  
The symmetric Zhang $SO(5)$ model 
\cite{zha97}   
predicts no charge gap at half filling. 
To recover 
Mott insulator phases at half-filling in the Zhang model it is normal to 
impose 
a no-double-occupancy rule by Gutzwiller projection. This breaks $SO(5)$ symmetry, 
but lattice calculations 
and schematic arguments 
suggest that many $SO(5)$ features might survive in such a projected $SO(5)$ model 
\cite{zha99,zac00,arr00,dor02}. 
 
We have proposed a 
unified description of high temperature 
superconductivity and antiferromagnetism based on a $U(4) \supset SU(4)$ 
dynamical symmetry that has analytical solutions in three 
symmetry limits \cite{gui99,wu03}.  
The $SO(4)$ limit of the $SU(4)$ model 
corresponds to an antiferromagnetic phase, the 
$SU(2)$ limit to a $d$-wave superconducting phase,  
and the $SO(5)$ limit  
to a critical symmetry interpolating between the antiferromagnetic and 
superconducting phases.  
 
Although the methodology of the $SU(4)$ model differs substantially  
from that of the Zhang model,  
its $SO(5)$ limit  
represents the Zhang $SO(5)$ algebra subject to 
constraints implied by embedding $SO(5)$ in the 
larger algebra $SU(4)$. 
In this Letter we address the physical 
understanding of why $SU(4)$ should play a crucial 
role in high temperature superconductivity, how  
no-double-occupancy and Mott insulator properties lie at the basis of this 
understanding, and provide a microscopic understanding of the projected 
$SO(5)$ model. 
 
The $U(4) \supset SU(4)$ model has 16 symmetry generators: 
\begin{eqnarray}  
p_{12}^\dagger&=&\sum_k g(k) c_{k\uparrow}^\dagger 
c_{-k\downarrow}^\dagger 
\qquad p_{12}=\sum_k g^*(k) c_{-k\downarrow} c_{k\uparrow} \nonumber 
\\  
q_{ij}^\dagger &=& \sum_k g(k) c_{k+Q,i}^\dagger c_{-k,j}^\dagger 
\qquad q_{ij} = (q_{ij}^\dagger)^\dagger \label{eq1} 
\\  
Q_{ij} &=& \sum_k c_{k+Q,i}^\dagger c_{k,j} \qquad S_{ij} = \sum_k 
c_{k,i}^\dagger c_{k,j} - \tfrac12 \Omega \delta_{ij}  \nonumber 
\end{eqnarray}  
where $c_{k,i}^\dagger$ creates an electron of momentum $k$ and  
spin projection $i,j= 1 {\rm\ or\ }2 \equiv \ \uparrow$ or 
$\downarrow$, $Q=(\pi,\pi,\pi)$ is an AF ordering vector, $\Omega/2$ is the 
electron-pair degeneracy, and  $g(k)$ is the $d$-wave form 
factor 
\begin{equation} 
g(k)=(\cos k_x -\cos k_y) \approx {\rm sgn} (\cos k_x -\cos k_y). 
\label{formfactor} 
\end{equation} 
By forming new 
linear combinations, (\ref{eq1}) may be replaced by  
operators having more direct physical meaning: 
\begin{eqnarray}  
Q_+&=&Q_{11}+Q_{22} = \sum_k (c_{k+Q\uparrow}^\dagger 
c_{k\uparrow} + c_{k+Q\downarrow}^\dagger c_{k\downarrow}) \nonumber 
\\ 
\vec S &=& \left( \frac{S_{12}+S_{21}}{2}, 
                \ -i  \frac {S_{12}-S_{21}}{2}, 
                \ \frac {S_{11}-S_{22}}{2} \right) \nonumber 
\\ 
\vec {\cal Q} &=& \left(\frac{Q_{12}+Q_{21}}{2},\ -i \frac{Q_{12}-Q_{21}}{2}, 
\ \frac{Q_{11}-Q_{22}}{2} \right) 
\label{operatorset} 
\\ 
\vec \pi^\dagger &=& \left( i\frac {q_{11}^\dagger - q_{22}^\dagger}2, \ 
\frac{q_{11}^\dagger + q_{22}^\dagger}2, 
\ -i\frac {q_{12}^\dagger + q_{21}^\dagger}2 \right) \nonumber 
\\ 
\vec \pi&=&(\vec \pi^\dagger)^\dagger, 
\ D^\dagger = p^\dagger_{12}, 
\  D = p_{12}, 
\  M=\tfrac12 (S_{11}+S_{22}) \nonumber 
\end{eqnarray}  
where $Q_+$  is  associated 
with charge density waves, 
$\vec S$ is the spin operator, $\vec {\cal Q}$ is the staggered magnetization,  
$D^\dagger$ ($D$) is the creation (annihilation) operator of spin-singlet $d$-wave 
pairs, $\vec \pi^\dagger$ ($\vec \pi$) are associated with spin-triplet 
pairs, and $2M=n-\Omega$ is related to the number (charge) operator $n$. 
The representation space of the 
$SU(4)$ model is  built by the coherent $D$ and 
$\pi$ pairs: 
\begin{equation} 
\ket{n_x n_y n_z n_s} =  (\pi_x^\dagger)^{n_x} (\pi_y^\dagger)^{n_y} 
(\pi_z^\dagger)^{n_z} (D^\dagger)^{n_s} 
\ket{0}. 
\label{collsubspace} 
\end{equation} 
The operator $Q_+$  
commutes with all generators 
and will be ignored in this discussion.  
Thus the most general effective Hamiltonian in the symmetry-dictated  
truncated space is a linear combination of all scalar 
products constructed from the remaining 15 $SU(4)$ generators.  Assuming 
only one and two-body interactions,  
\begin{eqnarray} 
  H = \hat{n}\varepsilon &-& v \hat{n}^2 
-G_0 D^\dag D  -G_1\vec{\pi}^\dag\cdot\vec{\pi}\nonumber \\            
&-&\chi\vec{\cal Q}\cdot\vec{\cal Q}+g\vec{S}\cdot\vec{S},          
\label{eqh1} 
\end{eqnarray} 
where $\varepsilon$, $v$, $G_0$, $G_1$, $\chi$, and $g$  
define effective microscopic strengths of single-particle and interaction  
terms.  
 
The operator set (\ref{operatorset}) is closed under $U(4)\supset 
U(1)\times SU(4)$ symmetry  (hereafter termed $SU(4)$) 
only if the approximation in 
(\ref{formfactor}) holds.  The physics  
implied in this approximation becomes more 
transparent if we transform (\ref{eq1}) to 
coordinate space (using the exact form of Eq.\ (\ref{formfactor})):   
\begin{eqnarray}  
p_{12}^\dagger&=&\sum_r  c_{{\bf r}\uparrow}^\dagger 
c^\dagger_{\bar{\bf r}\downarrow}\   
\hspace{36pt} p_{12}=\sum_r  
c_{\bar{\bf r}\downarrow}c_{{\bf r}\uparrow} \ \nonumber \\  
q_{ij}^\dagger\hspace{3pt}&=&\sum_r(-)^r c_{{\bf r},i}^\dagger 
c^\dagger_{\bar{\bf r},j}\   
\hspace{14pt} q_{ij}\hspace{3pt}=\sum_r (-)^r 
c_{\bar{\bf r},j}c_{{\bf r},i} \ \label{rspace} \\  
{Q}_{ij}&=&\sum_r(-)^r  c_{{\bf r},i}^\dagger 
c_{{\bf r},j}\   
\hspace{14pt} S_{ij}\hspace{2pt}=\sum_r c_{{\bf r},i}^\dagger 
c_{{\bf r},j} - \frac12\Omega\delta_{ij}\nonumber\  
\end{eqnarray}  
where $c^\dagger_{{\bf r},i}$ ($c_{{\bf r},i}$) creates (annihilates) an electron 
of spin $i$ located at ${\bf r}$ and $c^\dagger_{\bar{{\bf r}},i}$ 
($c_{\bar{{\bf r}},i}$) creates (annihilates) an electron of spin $i$ at the four 
neighboring sites, ${\bf r}\pm{\bf a}$ and ${\bf r}\pm{\bf b}$, with equal 
probabilities (${\bf a}$ and ${\bf b}$ are lattice constants in ${\bf x}$ 
and ${\bf y}$ directions, respectively), 
\begin{equation} 
c^\dagger_{\bar{\bf r},i}=\frac 12 
\left(c^\dagger_{{\bf r} 
+{\bf a},i}+c^\dagger_{{\bf r}-{\bf a},i}  
-c^\dagger_{{\bf r}+{\bf b},i} 
-c^\dagger_{{\bf r}-{\bf b},i}\right) . 
\label{cbar} 
\end{equation}  
The factor $(-)^r$ in Eq.\ (\ref{rspace}) is $(-)^{n_x+n_y}$  and 
($n_x,n_y$) are the coordinates of a lattice  
site on the copper oxide plane, 
${\bf r}=n_x{\bf a}+n_y {\bf b}$,  
which is positive for even sites ($n_x+n_y=$ even) 
and negative for the odd sites ($n_x+n_y=$ odd). This factor originates from the 
assumption $e^{iQ\cdot{\bf r}}\approx (-)^r$ and implies Mott 
insulator properties:  the 
electrons are localized at lattice sites with small overlap 
between orbitals of electrons on neighboring lattice sites. 
  
\epsfigbox{fig1}{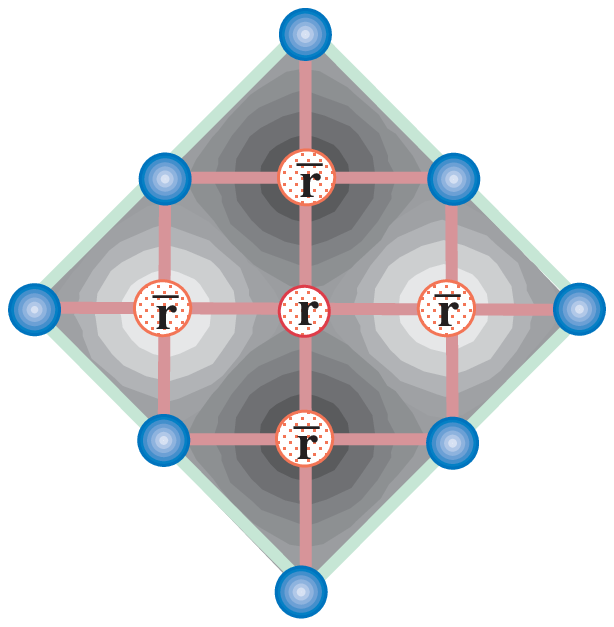}{0pt}{0pt} 
{(Color online) A schematic hole pair.  Fuzzy balls are  
sites where electron holes form a pair:   one 
hole at {\bf r}, the other with equal probability 
(1/4) at the four neighboring sites  
(${\bf \bar{r}=r\pm a}$ and ${\bf r\pm b}$). 
Balls connected by bright lines are  
sites where the presence of a hole would imply double occupancy. } 
 
From the coordinate representation (\ref{rspace}) we see that  
spin-singlet and spin-triplet pairs are formed by 
holes on adjacent sites.  
Fig.\ 1 illustrates the spatial structure of a hole pair: 
if one hole 
is at ${\bf r}$, the other hole occupies the four adjacent 
sites (${\bf r}\pm{\bf a}$ and ${\bf r}\pm{\bf b}$) with equal probability. 
The summation over ${\bf r}$ in the pair creation 
(annihilation) operators indicates that such pairs are highly coherent. 
It also can be seen that  
\begin{eqnarray}  
\hat{n}&=&\hat{n}^{(e)}+\hat{n}^{(o)}\   
\hspace{14pt} Q_+=\hat{n}^{(e)}-\hat{n}^{(o)} \\  
S_{ij}&=&S_{ij}^{(e)}+S_{ij}^{(o)}\   
\hspace{12pt} {Q}_{ij}=S_{ij}^{(e)}-S_{ij}^{(o)}  
\end{eqnarray}  
with $\hat{n}^{(e)}$ ($\hat{n}^{(o)}$) and  $S_{ij}^{(e)}$ ($S_{ij}^{(o)}$) the  
total electron number and spin operators at even (odd) sites, 
respectively: 
\begin{eqnarray}  
\hat{n}^{(e)}&=&\sum_{i,r= {\rm even}}  c_{{\bf r},i}^\dagger c_{{\bf r},i}\  
\hspace{10pt}\hat{n}^{(o)}=\sum_{i,r= {\rm odd}} c_{{\bf r},i}^\dagger c_{{\bf 
r},i}\  
\label{eqn} \\ 
S_{ij}^{(e)}&=&\sum_{r= {\rm even}}  c_{{\bf r},i}^\dagger c_{{\bf r},j}\  
\hspace{10pt}S_{ij}^{(o)}=\sum_{r= {\rm odd}} c_{{\bf r},i}^\dagger c_{{\bf r},j}\ . 
\label{eqs}  
\end{eqnarray}  
Thus $Q_+$ and $\vec{Q}$  
represent the differences in total charge and spin between even and odd 
sites, respectively.  
 
However, (\ref{rspace}) does not close under commutation 
unless 
\begin{equation} 
\left \{ c_{\bar{\bf r}',i}\ , c^\dagger_{\bar{\bf r},j} \right\}= 
\delta_{\bf r'r}\delta_{ij}\ ,\hspace{14pt} 
\left \{ c_{\bar{\bf r}',i}\ , c_{\bar{\bf r},j} \right\}=0\  
\label{cbarcom1} 
\end{equation} 
(that is, $c_{\bar{\bf r},i}^\dagger$ ($c_{\bar{\bf r},i}$) is  
a basis for particles occupying sites adjacent to {\bf r}).  
This separates  
lattice sites into categories {\em A} and {\em B}: if 
$r$ = even are {\em A} sites (with operators $c_{{\bf r},i}^\dagger$ 
($c_{{\bf r},i}$)), $r$ = odd are the {\em B} sites (operators 
$c_{\bar{\bf r},i}^\dagger$ ($c_{\bar{\bf r},i}$)), or vice versa.  
Then (\ref{cbarcom1}) permits (\ref{rspace}) to be written as  
\begin{eqnarray}  
p_{12}^\dagger&=&\hspace{10pt} \sum_{{\bf r}\in A} \left( c_{{\bf 
r}\uparrow}^\dagger c^\dagger_{\bar{\bf r}\downarrow}\ -\ c_{{\bf 
r}\downarrow}^\dagger c^\dagger_{\bar{\bf r}\uparrow}\right) \nonumber \\  
q_{ij}^\dagger\hspace{3pt}&=&\pm\sum_{{\bf r}\in A} \left( c_{{\bf 
r},i}^\dagger c^\dagger_{\bar{\bf r},j}+c_{{\bf r},j}^\dagger 
c^\dagger_{\bar{\bf r},i}\right) \nonumber \\ 
S_{ij}\hspace{2pt}&=&\hspace{10pt}\sum_{{\bf r}\in A} \left( c_{{\bf r},i}^\dagger 
c_{{\bf r},j}-c_{\bar{\bf r},j} c_{\bar{\bf r},i}^\dagger\right) 
\label{rspace2}\\   
\tilde{Q}_{ij}&=&\pm\sum_{{\bf r}\in A} \left( c_{{\bf r},i}^\dagger c_{{\bf 
r},j}+c_{\bar{\bf r},j}c_{\bar{\bf r},i}^\dagger\right)\nonumber\\  
p_{12}&=&(p_{12}^\dagger)^\dagger \qquad  
q_{ij}\hspace{3pt}=(q_{ij}^\dagger)^\dagger  
\nonumber 
\end{eqnarray} 
with 
$\tilde{Q}_{ij}\equiv Q_{ij}+\tfrac{\Omega}{2}\delta_{ij}$, where the $\pm$ sign 
is $+$ ($-$) if {\em A} is chosen to be even (odd) sites.    
(Whether {\em A} sites are taken to be even or odd is a labeling choice and  
does not influence the physics.) Then by explicit 
commutation the operators (\ref{rspace2}) close an $SU(4)$ algebra. 
But by Eq.\ (\ref{cbar}),  
\begin{eqnarray} 
&&\hspace{-24pt}\left \{c_{\bar{\bf r}',i}\ , c^\dagger_{\bar{\bf r},j} \right\}= 
\delta_{\bf r'r}\delta_{ij}+\delta_{ij}\frac14 
\sum_t g(t)\ \delta_{\bf r',r+t}\label{cbarcom2}\\ 
&&\hspace{-24pt}\mbox{with }\left\{\begin{array}{l} 
g(t)=+1\mbox{ for }{\bf t=\pm 2a, \pm 2b},\\ 
g(t)=-1\mbox{ for }{\bf t=+a \pm b,-a\pm b} \label{eq15b} \end{array}\right. 
\end{eqnarray} 
and  
(\ref{cbarcom1}) is generally {\em not} satisfied 
unless the second term on the right side of 
(\ref{cbarcom2}) can be ignored. This term vanishes if  a 
constraint is imposed that whenever there is an electron pair 
$c^\dagger_{{\bf r}i}c^\dagger_{\bar{\bf r}j}$ at ${\bf r}$ (see Fig.\ 1),  
no pair is permitted at 
${\bf r'}={\bf r+t}$, leaving nothing  
to be annihilated by $c_{\bar{\bf r}'i}$.  
This is a no-double-occupancy constraint because without it  
there is a finite amplitude for double site occupancy. For 
instance, if one pair is at ${\bf r'}={\bf r}+2{\bf a}$ 
and a second pair at 
${\bf r}$,  the probability is  
$\tfrac{1}{16}$ for two electrons to be located at ${\bf r}+{\bf 
a}$ (see Fig.\ 1). We conclude that 
{\em closure of the $SU(4)$ algebra 
is a direct consequence of no double occupancy  
in the copper oxide 
conducting plane.} 
 
Additional insight follows from observing that 
the 
validity of (\ref{cbarcom1}) actually follows 
from the more general requirement that  
no pairs overlap,  
a consistency condition 
ensuring that the pair space and 
the pairing correlations be well defined, is sufficient to 
satisfy (\ref{cbarcom1}).  
The no-pair-overlap constraint implies naturally 
that if a pair is centered at ${\bf r}$,  no pair may be located 
at ${\bf r'}={\bf r+t}$  with ${\bf t}$ given in (\ref{eq15b}),  
and thus Eq.\ (\ref{cbarcom1}) holds. 
 
For an $N$-dimensional basis the 
minimum closed  
algebra is $SO(2N)$ if all bilinear particle--hole and 
pair operators are  
taken as generators. 
The simplest basis for cuprates may be regarded as 
4-dimensional since 
electrons can exist only in four basic states,  
on {\em A}-sites 
or {\em B}-sites, with spin up or down.  
Thus, absent further constraints,  
the minimum Lie algebra for the 
set of generators that can describe high $T_c$ superconductivity and 
antiferromagnetism simultaneously in a cuprate system is $SO(8)$ and not 
$SU(4)$.  
The 28 generators of $SO(8)$ are the 16 operators in  (\ref{rspace2}) plus the 
12 operators 
\begin{eqnarray}  
\bar{p}_{12}^\dagger&=&\hspace{10pt}\sum_{{\bf r}\in A} \left( c_{{\bf 
r}\uparrow}^\dagger c^\dagger_{{\bf r}\downarrow}\ -\ c^\dagger_{{\bf 
\bar{r}}\downarrow}c_{ \bf 
\bar{r}\uparrow}^\dagger\right) \nonumber \\  
\bar{q}_{12}^\dagger\hspace{3pt}&=&\pm\sum_{{\bf r}\in A} \left( c_{{\bf 
r}\uparrow}^\dagger c^\dagger_{{\bf r}\downarrow}\ + \ c^\dagger_{{\bf 
\bar{r}}\downarrow}c_{ \bf 
\bar{r}\uparrow}^\dagger\right) \nonumber \\    
\bar{S}_{ij}\hspace{2pt}&=&\hspace{10pt}\sum_{{\bf r}\in A}  \left( c_{{\bf 
r},i}^\dagger c_{{\bf \bar{r}},j}-c_{{\bf r},j}c_{{\bf 
\bar{r}},i}^\dagger\right) 
\label{so8}\\   
\bar{Q}_{ij}&=&\pm\sum_{{\bf r}\in A} \left( c_{{\bf r},i}^\dagger c_{{\bf 
\bar{r}},j}+c_{{\bf r},j}c_{{\bf \bar{r}},i}^\dagger\right)\nonumber\\ 
\bar{p}_{12}&=&(\bar{p}_{12}^\dagger)^\dagger \qquad 
\bar{q}_{12}\hspace{3pt}=(\bar{q}_{12}^\dagger)^\dagger 
\nonumber 
\end{eqnarray} 
($\pm$ depends on the even--odd choice for {\em A} sites; see (\ref{rspace2})).   
 
Equation (\ref{so8}) 
contains two kinds of new (spin-singlet) pairs created by 
$\bar{p}_{12}^\dagger$ and $\bar{q}_{12}^\dagger$, 
which may be termed $S$ and $S^*$ pairs, respectively.  In both the two 
electrons (holes) occupy  the same site, with equal probability to appear 
anywhere in the lattice coherently. 
The  $S^*$ pairs differ from $S$ 
pairs in their phases. 
The operators 
$\bar{S}_{ij}$ are the hopping operators with and without 
spin flip, and $\bar{Q}_{ij}$ is the staggering of the hopping. 
These operators change $D$ and $\pi$ pairs into $S$ and $S^*$, or vice versa. 
 
The $SO(8)$ algebra reduces to $SU(4)$ if the 
$S$ and $S^*$ pairs may be neglected,   
which occurs if we assume on-site Coulomb repulsion 
pushing the $S$ and $S^*$ pairs to sufficiently high energy.  
Thus,  restriction to no double occupancy effectively 
allows the operators in 
Eq.\ (\ref{so8}) to be ignored 
and reduces $SO(8)$ to the subalgebra $SU(4)$. 
 
Therefore, the minimal Lie algebra  
that can describe antiferromagnetism and $d$-wave superconductivity in a  
cuprate system is in general 
$SO(8)$, but under the constraint of 
no double occupancy the symmetry effectively reduces to  
$SU(4)$.   
{\em The assumption of an $SU(4)$ symmetry in a cuprate system 
automatically implies the 
imposition of a no-double-occupancy constraint on the 
general $SO(8)$ symmetry in the copper--oxygen 
planes.}  
 
It is likely that the hopping  
operator $\bar{S}_{ij}$ in Eq. (\ref{so8}) 
is the source of the most important $SU(4)$ symmetry breaking 
terms.  It breaks $SU(4)$ but is a generator of 
$SO(8)$, so    
this perturbation may be taken into account by an  
extension from $SU(4)$ to $SO(8)$.   
However, we may expect 
the  no-double-occupancy rule and thus the $SU(4)$  
symmetry to be a good initial approximation. 
 
\epsfigbox{fig2}{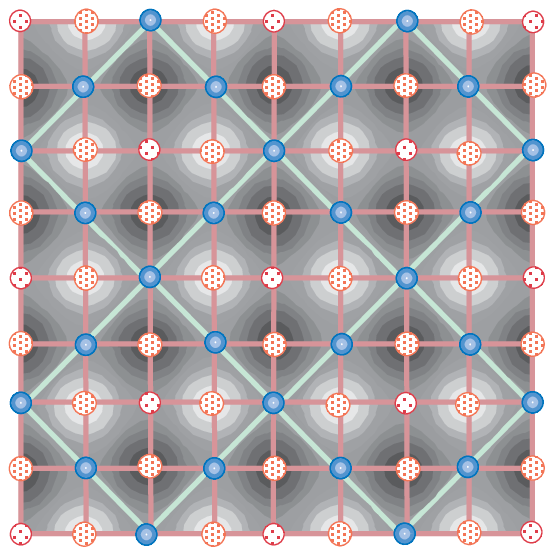}{0pt}{0pt} 
{(Color online) Schematic    
spatial distribution for maximal hole pair occupation. 
Fuzzy balls are lattice sites where the electron holes form 
pairs. Balls connected by bright lines indicate 
sites where the presence of a hole would lead to double occupancy. } 
 
The implicit $SU(4)$ occupancy constraint  
dictates an upper limit for the doping fraction in $SU(4)$-conserving states. 
Fig.\ 2 illustrates   
the spatial distribution when the 
number of hole pairs is maximal.  
By counting,  
the maximum number of holes 
is 
$\Omega=\Omega_e/4$, where $\Omega_e$ is the total number of lattice 
sites. Thus   
the 
largest doping fraction   
preserving $SU(4)$ symmetry is $P_f=\Omega/\Omega_e= \tfrac14$.  
The empirical  
maximum doping fraction ($0.23\sim 0.27$) 
for cuprate superconductivity may then be interpreted as a direct consequence of 
$SU(4)$ symmetry. 
   
The preceding discussion implies that: 
(1)~The physical origin of cuprate $SU(4)$ 
symmetry is proximity of antiferromagnetism and $d$-wave pairing, 
coupled with suppression of double occupancy by 
on-site Coulomb repulsion. 
(2)~Two important facts in cuprates, that normal states are  
Mott insulators and 
that superconductivity exists only in a  
narrow doping range ($P < 0.27$), are 
direct consequences of an $SU(4)$ dynamical symmetry.  
 
Superconductivity in cuprates is a specific example of  
what we shall term 
{\em non-abelian superconductivity,}   
which differs 
from conventional superconductivity in the 
richness of pair structure for condensed states and 
in the appearance of competing sources of long-range order. 
The key issues for $SU(4)$ non-abelian superconductivity 
are that coherent pairs are formed by  
holes on adjacent sites  
so that both singlet and triplet states  
contribute, and that alternative long-range order (antiferromagnetism) 
enters on an equal footing 
with superconductivity.  
 
In contrast to BCS superconductivity, which is 
described by a single dynamical 
symmetry chain having only abelian subgroups  
($SU(2) \supset U(1)$), the minimal symmetry consistent 
with cuprate data is $SU(4)$, which has a much richer  
structure (three dynamical symmetries having non-abelian subgroups and 
differing fundamentally in their properties). 
We propose that the differences in observational characteristics for 
these two types of superconductivity 
originate in this difference in 
dynamical symmetry structure and in 
non-abelian superconductivity resulting from 
electron--electron interactions instead of electron--phonon interactions. 
 
The primal role of $SU(4)$ symmetry in non-abelian 
cuprate superconductivity suggests  
that any pairing structure leading to the $SU(4)$ algebra 
entails dynamics similar to that of cuprates. 
Therefore, $d$-wave symmetry of the 
pairs need not be critical to non-abelian superconductivity in general 
and $SU(4)$  
superconductivity in particular.   Pairs  
with any internal symmetry (extended $s$-wave,  
$p$-wave, mixed symmetry, \ldots)  
could exhibit $SU(4)$ superconductivity if the 
no-double-occupancy constraint is valid and correlations can form  
adjacent-site pairs. 
Generally,  
$c^\dagger_{\bar{\bf r},i}$ may be defined as 
\begin{equation} 
c^\dagger_{\bar{\bf r},i}=\sum_{\bf t} g({\bf t}) 
c^\dagger_{{\bf r}+{\bf t},i}\ \qquad 
\sum_{\bf t} \left | g({\bf t})\right |^2=1 
\label{gt} 
\end{equation}  
where {\bf t} is a few finite lattice displacements of ${\bf r}$ and 
$g({\bf t})$ is the form factor.  Different forms of $g({\bf t})$ reflect 
different internal symmetries of the pairs, but they 
all satisfy the condition (\ref{cbarcom1}) under  
no double occupancy and thus preserve the $SU(4)$ algebra and the 
general Hamiltonians implied by its dynamical symmetry chains.  The structure 
(\ref{cbar}) of the  $d$-wave pairs is only a  
special case of (\ref{gt}) with  
${\bf t}=\pm a , \pm b$ and 
$g(\pm{\bf a})=\tfrac12$ and $g(\pm{\bf b})=-\tfrac12$. 
 
In summary, we have shown that  
$SU(4)$ is the minimal symmetry accommodating superconductivity, 
antiferromagnetism, and a no-double-occupancy constraint in 
cuprate systems, and that $SU(4)$ symmetry implies a maximum doping fraction 
of $\tfrac14$ in the cuprates, by symmetry alone. 
Because the Zhang $SO(5)$ algebra  
is a subalgebra of  
$SU(4)$,   
these results indicate that {\em closure of the Zhang algebra also implies 
no double occupancy.} Why then does the $SO(5)$ model require Gutzwiller 
projection?   
The work presented here suggests that the projection requirement 
arises from assumptions inconsistent with the underlying $SO(5)$ algebra 
in the Zhang effective Lagrangian formulation,  
where five of the 
$SU(4)$ generators ($D^\dagger, D, \vec{\cal Q}$) are treated as 
order parameters forming a superspin vector. Thus one cannot 
apply algebraic constraints to them through the commutators. This is 
most easily seen if not only the $SO(5)$ generators but also the elements of 
the Zhang superspin vector are treated as operators 
rather than order parameters, thereby enlarging the algebra to 
$SU(4)$. Within the $SU(4)$ framework, 
there is no need for projection. As demonstrated in Refs. \cite{gui99, 
wu03}, within the parent $SU(4)$ group 
antiferromagnetism and superconductivity 
are described by {\em different} dynamical symmetries 
($SO(4)$ and $SU(2)$, respectively), and $SO(5)$ is a 
critical dynamical symmetry that interpolates between $SO(4)$ and $SU(2)$.  
In the $SU(4)$ model it is the $SO(4)$ symmetry, not the 
$SO(5)$ symmetry, that naturally describes undoped states, and the 
spectrum for {\em unbroken} $SO(4)$ dynamical  
symmetry is intrinsically antiferromagnetic  
with gapped charge excitations. 
 
Our results imply some important consequences  
of attributing cuprate superconductor 
behavior to an $SU(4)$ algebra that follow directly  
from symmetry, independent of details: (1)~Normal states 
are antiferromagnetic Mott insulators.  
(2)~Hole doping of normal compounds leads first to $SO(5)$ 
fluctuations in both antiferromagnetic and superconducting order (implying 
phases that may exhibit  
spin glass or stripe character), and then to $SU(4)$  
non-abelian superconductivity. 
(3)~$SU(4)$ superconductivity 
is strongly suppressed 
for doping fractions exceeding $\tfrac14$.  
(4)~Symmetry breaking resulting from violation of the no-double-occupancy  
constraint may be described by a parent $SO(8)$ algebra where 
terms that break $SU(4)$ may still 
respect $SO(8)$ symmetry. 
(5)~$SU(4)$ symmetry, not $d$-wave pairing {\em per se}, is the ultimate cause of 
cuprate behavior, implying that systems could exist having non-$d$  
pairing but cuprate-like dynamics.   
The first three consequences  
are postdictions 
in strong accord with existing data.  
The fourth is 
a prediction that  
may be tested through detailed applications of the $SU(4)$ model to data. 
The final prediction may be tested  
by searching experimentally for  
compounds  
having pairing 
structures other than $d_{x^2-y^2}$ 
that satisfy the $SU(4)$ algebra.  
 
We thank Aditya Joshi for discussions and for careful checking of the 
$SU(4)$ commutation relations. 
 
\baselineskip = 14pt 
\bibliographystyle{unsrt}

\end{document}